\documentclass{elsart}

\usepackage{amssymb}
\usepackage{graphics}
\usepackage{graphicx}
\usepackage[sectionbib,square]{natbib}

\begin{document}
\bibliographystyle{plainnat}
\begin{frontmatter}

\title{Evolving learning rules and emergence of cooperation in spatial Prisoner's Dilemma}
\author{Luis G. Moyano$^{\rm a}$ and Angel S\'anchez$^{\rm a,b,c}$}
\address[1]{Grupo Interdisciplinar de Sistemas Complejos (GISC), Departamento de Matem\'aticas,
Universidad Carlos III de Madrid, 28911 Legan\'es, Madrid, Spain}
\address[2]{Instituto de Ciencias Matem\'aticas CSIC-UAM-UC3M-UCM, 28006 Madrid, Spain}
\address[3]{Instituto de Biocomputaci\'on y F\'{\i}sica de Sistemas Complejos (BIFI), Universidad
de Zaragoza, 50009 Zaragoza, Spain}

\begin{abstract}
In the evolutionary Prisoner's Dilemma (PD) game, agents play with each other and
update their strategies in every generation according to some microscopic dynamical rule.
In its spatial version, agents do not play with every other but, instead, interact only with
their neighbors, thus mimicking the existing of a social or contact network that defines who
interacts with whom. In this work, we explore evolutionary, spatial PD systems consisting
of two types of agents, each with a certain update (reproduction, learning) rule.
We investigate two different
scenarios: in the first case, update rules remain fixed for the entire evolution of the system;
in the second case, agents update both strategy and update rule in every generation.
We show that in a well-mixed population the evolutionary outcome is always full defection.
We subsequently focus on two-strategy competition with nearest-neighbor interactions on the contact
network and synchronized update of strategies.
Our results show that, for an important range of the parameters of the game, the final
state of the system is largely different from that arising from the usual setup of a single,
fixed dynamical rule. Furthermore, the results are also very different if update rules are
fixed or evolve with the strategies. In these respect, we have studied representative
update rules, finding that some of them may become extinct while others prevail.
We describe the new and rich variety of final outcomes that arise from this co-evolutionary
dynamics. We include examples of other neighborhoods and asynchronous updating that
confirm the robustness of our conclusions.
Our results pave the way to an evolutionary rationale for modelling social
interactions through game theory with a preferred set of update rules.
\end{abstract}
\begin{keyword}
Game theory, evolution, prisoner's dilemma, learning, emergence of cooperation.
\end{keyword}

\end{frontmatter}

\section{Introduction}

The quest for the origins of the cooperative behavior observed in nature, in many different
species or at different biological levels, from molecules to individuals, is an exciting project
that has received much attention in the last decades \citep{Darwin,Axelrod,Maynard-Smith,%
Hammerstein}.
Evolutionary game theory has been one of the most successful frameworks to address this
issue in a quantitative manner \citep{Gintis,Nowakb,Pennisi} and by allowing a stylization of the main
strategic interactions and social dilemmas, has provided many insights into
the reasons of the emergence of cooperation.

A particularly fruitful line of research has focused on the interactions between two individuals,
trying to explain cooperative behaviors in a population with a bottom-up approach. In this
context, interactions are modelled by means of 2x2 games, such as the Prisoner's Dilemma
\citep{Rapoport} or the Hawk-Dove game \citep{Maynard-Viejo}. These games have proven
themselves relevant in situations arising in biochemistry \citep{Frick}, cooperation between
bacteria \citep{Crespi}, mutualistic interactions \citep{Kiers}, fish \citep{Dugatkin} and, of
course, human societies \citep{Kollock}.

Within the framework of evolutionary game theory, a number of explanations have been
proposed to understand the origin of cooperation \citep{Nowaka}. In this work, we focus on
one of them, namely the existence of a spatial structure, as such or as a representation of
a social network. Indeed, many studies have pointed out that certain types of spatial structure
foster cooperation in simple two-player symmetric games, beginning with the pioneering
work by \cite{Nowak-May}. Subsequent papers
\citep{Nowak-May2,Hauert-Viejo,Hauert,Santosa,Santosb,Raul,yo,bifi,Langer}
analysed different aspects of the emergence of cooperation in spatial games and as a
conclusion of this work the general idea that spatial structures supported cooperative
behaviors began to shape up.
For a comprehensive
summary of all the recent work on evolutionary game theory on graphs, see the review by
\cite{Szabo}.

Recently, some authors have questioned the generality of the above assertion, at least
as far as other games are concerned. Thus, \cite{Hauert-Doebeli} and \cite{Kaski} have
shown that spatial structure may decrease the cooperation level attained in the Hawk-Dove
game as compared to that observed in a well-mixed population. Following this result,
some researchers have looked in detail into the different reports and found that the
phenomenon of the emergence of cooperation, when truly existing, turns out to be
dependent on the microscopic update rules used in the simulation. In other words, it is
possible that within the same game, played in the same spatial structure, cooperation
arises or not depending on the way the players change their strategy during evolution.
This is the case, for instance, when playing Prisoner's Dilemma on a square lattice:
whereas unconditional imitation (see below for a description of this and the rest of
update rules studied in this paper) gives rise to cooperation \citep{Nowak-May}, replicator
dynamics leads to full defection \citep{Carlos}. Similar dependences of the results for
other update dyamics have been reported by \cite{joputa} for death-birth, birth-death
and imitation.

In view of this situation, in this paper we aim to going beyond the approach that has been
traditionally used in the study of spatial games. Specifically, we intend to provide an
evolutionary rationale for choosing a particular update rule in the implementation of spatial
models of cooperation. To this end, we will allow agents to update not only their strategies
but also the update rule itself. The outcome of these simulations will show whether or not all
update rules are equally likely to appear in a structured population and, if not, which ones
are evolutionarily selected. One can then compare this conclusion to the scenarios of
emergence of cooperation on networks and discuss the applicability of the different results
already known. Indeed, the fact that rules favoring cooperation were evolutionarily favored
would support the mechanism of network reciprocity as one of the most important ones for the
emergence of cooperation. On the contrary, if the competition among update rules leads to the
survival only of those that do not support cooperation, it would be difficult to argue that
networks promote cooperation.

For the present study,
we will be concerned only with the problem of one-shot or memoryless Prisoner's
Dilemma, as has been generally studied in the context of evolutionary game theory
on graphs; rules for deciding the action to be taken next on the basis of previous
ones, such as tit-for-tat, Pavlovian strategies, stochastic reactive strategies, etc.
Although it is possible to think of these strategies as C or D strategists with a
learning rule, their use of memory place them in a different class that and will
not be considered here. On the other hand, we here focus on a typical set of
local rules, as considered, for instance in \cite{Hauert-Viejo}; it is clear that this kind
of evolutionary competition may extend to many other update dynamics but an
exhaustive analysis of every possible rule is beyond the scope of the present research,
that intends only to assess the relevance of such an evolutionary process.

In this paper we address the co-evolution of strategies and update rules in a three-step
process. As a preliminary result, we discuss the case of well-mixed populations and show by an
example that in this situation including evolving update rules does not change the well-known
outcome, namely that defection prevails. Then we move to the case in which the population
interaction is governed by a lattice, as a specific example of social network in which it is
easier to interpret the results. In order to have a reference, a first step in our approach is
the comparison of mixed systems, consisting of agents with different (but fixed during
evolution) update rules, with the emergence of cooperation in pure single-dynamics systems.
Already at this stage we find differences between the two scenarios which are worth
describing; on the other hand, this is a necessary reference frame to understand the
subsequent results of co-evolution. Indeed, after this first step, we proceed to let update
rules evolve along with strategies. In this situation, we find new results that differ both
from the single-rule case and from the mixed-rule case. We will describe the results of our
simulations and in the conclusions we summarize our findings and discuss their implications
for the emergence of cooperation.

\section{Model}

Our model is based on the well-known Prisoner's Dilemma (PD) game \citep{Rapoport}.
An archetype in game theory, the PD game belongs to a general class of symmetric games that consist of two players or agents that confront, each with a definite strategy: either to cooperate or to defect with the opposite player.
Each combination of strategies between the players has an associated payoff, and hence there are four possible outcomes: if the player cooperates, the associated payoff she gets is $S$ if the other player chooses to defect, or $R>S$, if the other player reciprocates the cooperation.
On the other hand, if the first agent defects, her payoff is $P$, if the other also defects, or $T>P$, if the other cooperates.
The PD corresponds to any choice of payoffs ordered according to $T>R>P>S$.
It is customary to assume that  $2R > T+S$ to avoid that players take turns at defecting and win a larger payoff than the one they would have just cooperating. We will stick to this constraint although
in our case we deal with one-shot, non-repeated games because strategies and payoffs are
updated after every single game, and hence strictly speaking we need not impose this
condition.
In the rest of this work, we will adopt the commonly-used rescaled payoff $T=b$, $R=1$, and $P=S=0$ \citep{Nowak-May}.
This does not affect the general essence of the game, and reduces the study to just one free parameter, usually called the temptation parameter. Nevertheless, we checked the robustness of our
simulations by comparing some cases with the choice $S=-0.2$, finding very approximately
the same results.

In the evolutionary version of the PD game, $N$ agents play between them, and after every instance
of the game they decide whether to change strategies or not according to some microscopic update rule.
In a well-mixed situation, each player plays every other once and afterwards they proceed to
the update stage.
In our case, as we are interested in the rules that can promote cooperation in a spatially structured
population, the players are located at the nodes of a square lattice, where each agent is connected with her four closest  neighbors and plays the game only with those neighbors. The reason we have
chosen a square lattice is that it is a well studied model \citep{Nowak-May,Hauert-Viejo,Carlos,Langer} in
the single-update rule framework, and therefore we can compare our results to those previous
works. The sequence of steps for the simulation is as follows:
Each player is assigned, randomly and with equal probability, an initial strategy of cooperation (C) or defection (D), and a payoff, initially set to zero.
In each generation, all agents play PD once with each neighbour, with an associated payoff for each game, collecting a total final payoff for each player.
After each generation, all agents update their strategies simultaneously, according to a certain update rule (defined separately), that may depend on the agent's payoff and her neighbour's, and that defines the dynamics of the game.
Once defined, for all players, what the strategy for the next generation will be, all payoffs are reset to zero and the cycle starts over.

As we have said, the new ingredient of our model is twofold: two different update rules and the
possibility that the update rules themselves evolve.
Therefore, in our model players may be set to have rule A or rule B, where A and B stand for specific dynamical rules that will be explained below.
In this way, each player has its own individual dynamical rule.
Within this framework, we devised two possible alternatives.
In the first case, agents are assigned their dynamical rule, that remains fixed for the rest of the simulation.
In the second case, agents may change their dynamical rule during the game according to a simple criterion: an agent copies the dynamical rule from its neighbour whenever it copies its strategy.
This can be interpreted as a complete replacement of one agent for another agent's offspring, which may be convenient or useful in certain descriptions. This interpretation is suitable for both biological
and sociological situations, in this last case in terms of culture transmission and learning.
Furthermore, the possibility of variable update rules implies an evolution (and therefore a competition) of the rules themselves. If as a consequence of this evolution one rule, or a restricted set of rules,
are selected, the results on the emergence of cooperation on lattices will have to be revisited
again in the light of our findings.

We implement three of the most representative, and most commonly considered, dynamical rules: the unconditional imitation (UI, a.k.a. as imitate-the-best) rule, the Moran (MOR) rule and the replicator (REP) rule.
Unconditional imitation is a completely deterministic rule: at the end of each generation, an agent simply adopts the strategy of her neighbour with the highest payoff (i.e., the most successful one), given that this neighbour has a greater payoff. Note also that this rule checks the payoff of all the neighbors of the
agent whose strategy is to be updated. In this sense we will refer to this strategy as global (not to
be confused with global in terms of the whole lattice).
In the replicator dynamics, an agent randomly chooses another agent (in our case, one of its four neighbours) and, if the chosen agent has a higher payoff, the original one adopts that neighbour's strategy with a probability proportional to the difference of payoffs between the two. In this case,
we are faced with a local update rule, that does not look at all the updating agent's neighbourhood.
Another important remark is that, in our model, an agent having an imitation or replicator rule will never adopt another strategy (or rule) that performed worse.
Finally, in a Moran process, the agent, at the end of each generation, evaluates a set of probabilities, one for each neighbour and proportional to that neighbour's payoff.
Then the agent randomly selects a neighbour's strategy according to that set of probabilities.
In this case, there is a possibility that an agent adopts a strategy that performed worse in a previous round. On the other hand, this is again a global rule, in the same sense as we referred above to
the imitiation one.

\section{Well-mixed populations}
\label{sec:meanfield}

Before proceeding with the study of the evolutionary
competition of learning rules on lattices, it is important to
analyze the case in which every player plays against every
other one, i.e., a well-mixed population. The reason for the
need of such a study is twofold: on one hand, knowing the
behavior of a well-mixed population is necessary in order to
assess whether or not changing the scenario to a lattice has
any new effect; on the other hand, the well-mixed population
can be used to understand at least the initial stages of the
evolution on a lattice, when correlations arising because of
evolution are not yet important and the assumption that agents
meet an ``average'' agent can be made. We note that this
assumption is very common in the statistical physics literature
where it is referred to as the mean-field approach (see, e.g.,
\cite{Szabo} and references therein).

For the sake of simplicity, in what follows we will consider the case in which an initial fraction $x$ of imitator agents,
i.e., agents that learn through UI, compete with an initial fraction $1-x$ of replicator agents, agents using REP. The
other possible competitions can be treated in much the same way with similar results, therefore we use
this particular choice as an illustration. If initially a fraction $y$ of agents are cooperators (equally distributed
among UI and REP players), the four types of agents present in the population at time $t=0$ are given by
\begin{eqnarray}
\label{initial1}
f^0_{Ci}&=&xy,\\
\label{initial2}
f^0_{Di}&=&x(1-y),\\
\label{initial3}
f^0_{Cr}&=&(1-x)y,\\
\label{initial4}
f^0_{Dr}&=&(1-x)(1-y),
\end{eqnarray}
where the subindices $C$ or $D$ represent cooperators and defectors, respectively, and $i$ and $r$ refer
to imitators and replicators, also respectively.

Evolution begins by all agents playing the game with all the population. With our choice of payoffs, cooperators
receive a payoff $w_C=x$ and defectors receive a payoff $w_D=bx$. Let us now examine the evolutionary process
at the first step, beginning with cooperators:

{\em Imitator agents:} As $b>1$, we have $w_D>w_C$ and
therefore all imitator agents switch to defection at time $t=1$, and hence $f^1_{Ci}=0$; the newly created
defector will be an imitator with probability $x$ and a replicator with probability $(1-x)$.

{\em Replicator agents:} The fate of replicator
agents is  more complicated and in particular the evolution of their number in time
depends on the form chosen for the probability to
copy the other agent's strategy. If $p$ is the probability of a cooperator replicator to switch to defection
(the value of $p$ depends on the payoff difference, which is $b(1-x)$, and on the normalization), we have
the following scenario: The probability that a replicator chooses a cooperator to compare her strategy is
$y$ (the fraction of cooperators at $t=0$); in that case, the payoffs are the same and she will not change
her strategy and update rule. On the contrary, with probability $1-y$ she will compare with a defector,
and will turn into a defector herself with probability $p$. However, this defector will be an imitator with
probability $x$ and a replicator with probability $1-x$.

Finally, in view of the payoffs above, defectors never change, irrespective of their update rule.
Collecting all the different contributions, we have at $t=1$
\begin{eqnarray}
\label{uno1}
f^1_{Ci}&=&0,\\
\label{uno2}
f^1_{Di}&=&f^0_{Di}+xf^0_{Ci}+px(1-y)f^0_{Cr},\\
\label{uno3}
f^1_{Cr}&=&[y+(1-p)(1-y)]f^0_{Cr}=[1-p(1-y)]f^0_{Cr},\\
\label{uno4}
f^1_{Dr}&=&f^0_{Dr}+(1-x)f^0_{Ci}+p(1-x)(1-y)f^0_{Cr}.
\end{eqnarray}
From Eq.\ (\ref{uno3}) it is clear that the fraction of cooperator agents will steadily decrease; imitators disappear
at the first step and replicators will decrease exponentially [note that the coefficient in Eq.\ (\ref{uno3}) is smaller than 1]. As a consequence, asymptotically the population
will evolve to full defection. On the other hand, when all cooperator replicators
disappear, it can be shown straightforwardly that
they would have contributed to the two types of defectors simply proportionally to their initial fraction,
i.e.,
\begin{eqnarray}
\label{fin2}
f^{\infty }_{Di}&=&f^0_{Di}+x(f^0_{Ci}+f^0_{Cr})=x(1-y)+xy=x,\\
\label{fin4}
f^{\infty}_{Dr}&=&f^0_{Dr}+(1-x)(f^0_{Ci}+f^0_{Cr})=\\ \nonumber &=&(1-x)(1-y)+(1-x)y=(1-x).
\end{eqnarray}
This means that the effect of introducing a lattice as support of the game has indeed
non-trivial consequences because, as will be shown below, there is cooperation in a wide range of parameters,
and  the level of cooperation and the fractions of the different types of strategists and update rules can not be
predicted from the initial fractions in such an straightforward manner.

\section{Results on lattices}

To study our model, we performed a series of numerical experiments for both cases,
namely  with fixed and with variable update rules, for the same set of game parameters.
Each experiment consists in a population of $N=10^4$ agents, spatially arranged in a square lattice.
Generally speaking,
each agent is endowed initially with one of two available dynamical rules, A or B (where A and B stand for UI, MOR or REP), and one of two possible strategies, C or D.
To monitor the evolution of the system we will observe the frequency or density of a certain type of player, for example those that are cooperators or those with a certain update rule.
Thus, we define $f_{\scriptscriptstyle A}$ to be the number of agents with rule A divided by $N$, and $f^{\scriptscriptstyle C}$ the number of cooperator agents divided by $N$.
Every experiment is characterised by an initial density of cooperators $f^{\scriptscriptstyle C}(t=0)$ (or in a simpler notation, $f^{\scriptscriptstyle C}(0)$) and by an initial fraction of agents with rule A, $f_{\scriptscriptstyle A}(0)$.
In this way, $f^{\scriptscriptstyle C}$ will generally change. The homogeneous case (all agents
with the same update rule) is recovered by choosing $f_{\scriptscriptstyle A}(0)=1.$
To simplify the parameter space of our simulations, we only considered initial strategies  assigned randomly and with equal probability, i.e. $f^{\scriptscriptstyle C}(0)=0.5$ in all experiments, independent of the update rule considered.
All numerical simulations run for $T=10^4$ generations or time steps, for a given value of $b$.
Within this duration equilibration was achieved (often much earlier), i.e., the densities reached their
asymptotic values and remain there within small fluctuations.

\subsection{Fixed update rules}
\label{subsec:fixed}

We begin by looking at the simplest case of agents that can have two different update rules but they
cannot change them during evolution. In this case,
$f_{\scriptscriptstyle A}=f_{\scriptscriptstyle A}(0)$ and remains constant for all times.
It is useful to analyse the normalised fraction of cooperators of, say, rule A, which we will refer to
as $\mu_{\scriptscriptstyle A} = f_{\scriptscriptstyle A}^{\scriptscriptstyle C}/f_{\scriptscriptstyle A}$, where $0\leq\mu_{\scriptscriptstyle A}\leq1$, so we can better compare systems with different values if $f_{\scriptscriptstyle A}(0)$.

\begin{figure}[h!]
\centering
$\begin{array}{c@{\hspace{0.1in}}c}
\includegraphics[width=.5\textwidth,angle=0]{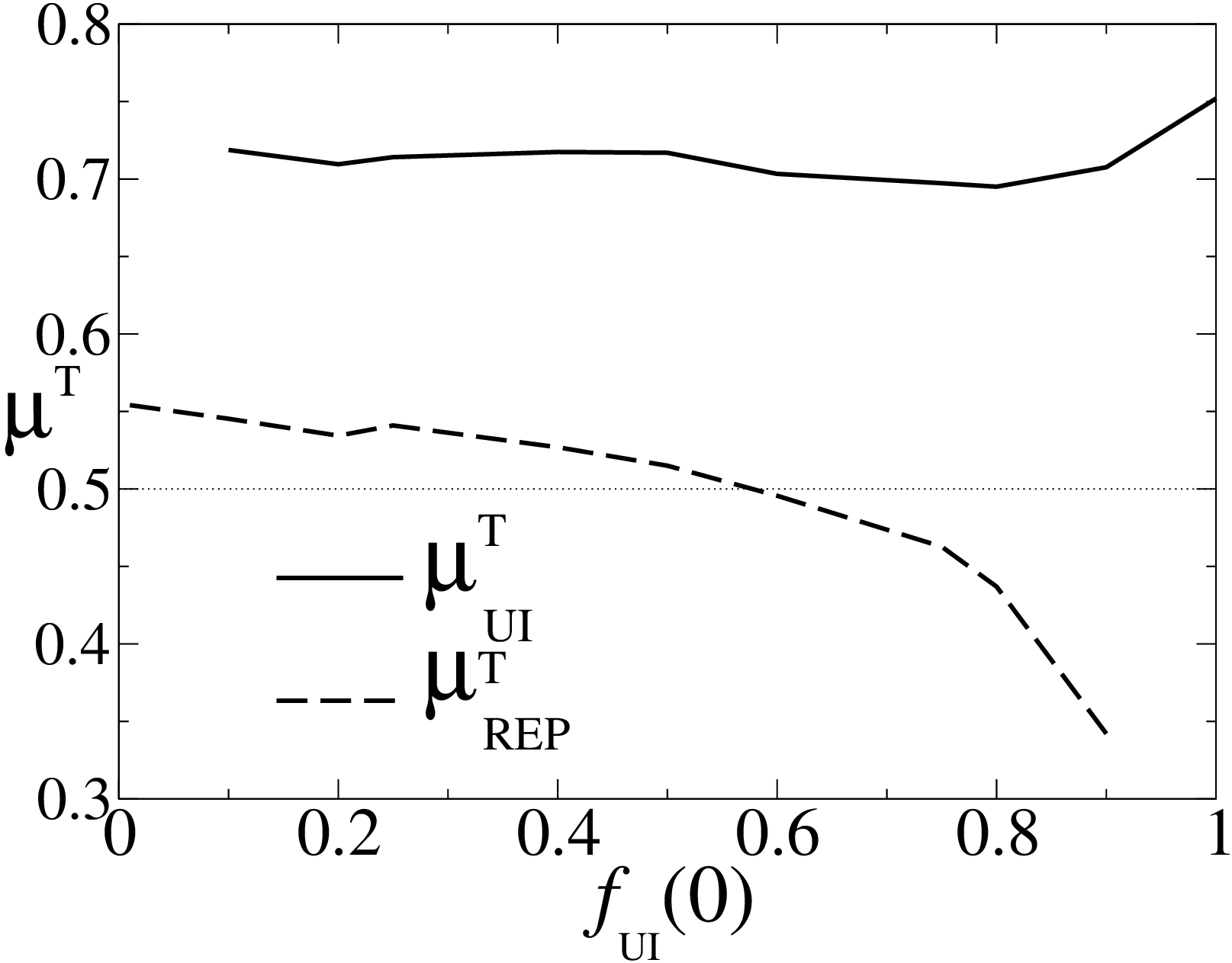}&
\includegraphics[width=.5\textwidth,angle=0]{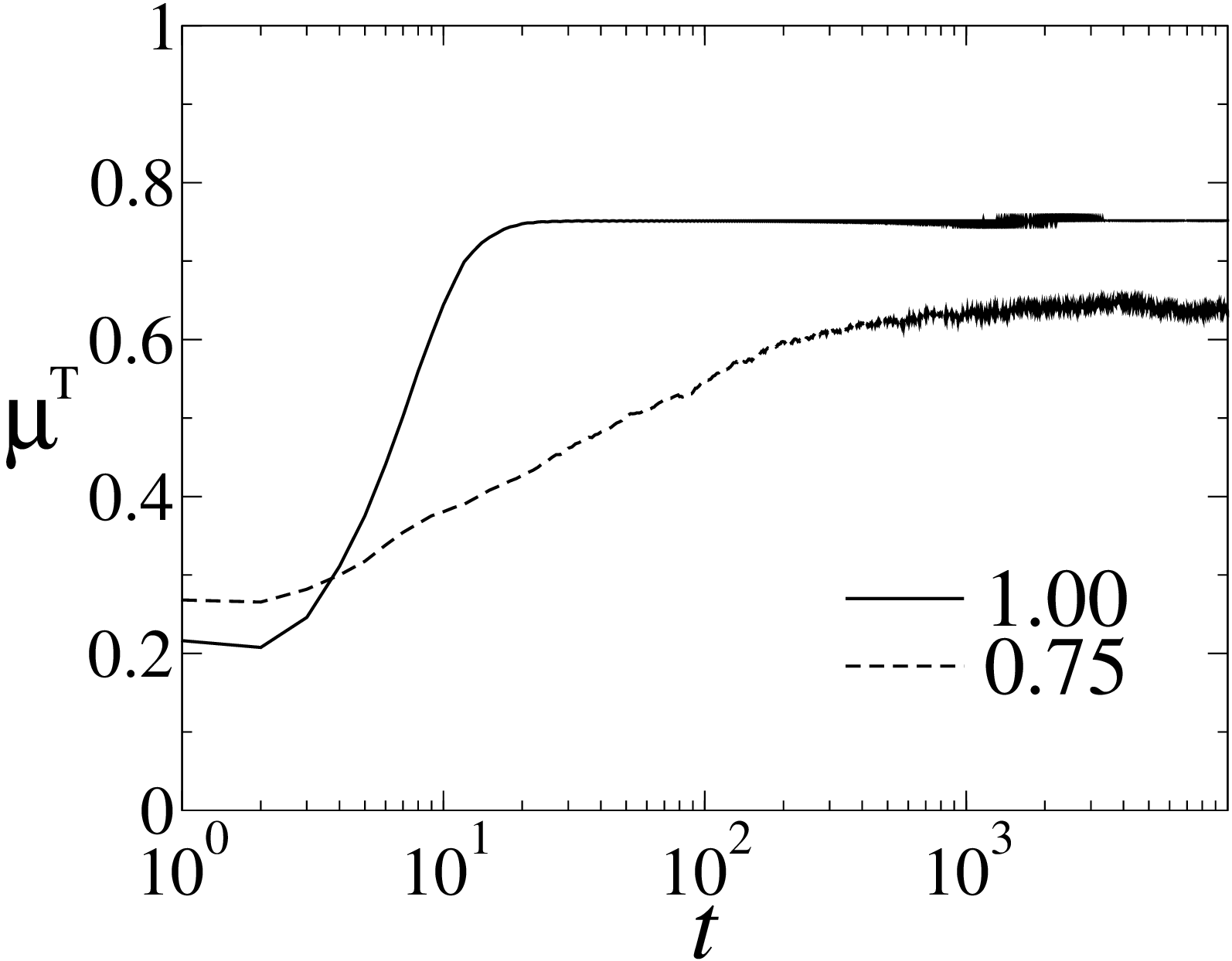}\\
\end{array}$
\caption{Left:
Value of $\mu^{\scriptscriptstyle T}_{\scriptscriptstyle UI}$ (solid) and $\mu^{\scriptscriptstyle T}_{\scriptscriptstyle REP}$ (dashed) at   $T=10^4$ as a function of the initial frequency of imitators agents $f_{\scriptscriptstyle UI}(0)$. Rules are fixed throughout the simulation.
Right: Time evolution of the normalised cooperator frequency $\mu_{\scriptscriptstyle UI} = f_{\scriptscriptstyle UI}^{
  C}/f_{\scriptscriptstyle UI}$ in the UI vs. REP game, for $b=1.05$ and two values of the initial frequency of imitators
$f_{\scriptscriptstyle UI}(0)$.
}
\label{fig1}
\end{figure}

A first finding arising from  our experiments is that the initial fraction of the populations, $f_{\scriptscriptstyle A}(0)$ may affect greatly the final cooperator outcome. Consider for
instance the left panel of Fig. \ref{fig1}, where we show an example of a system with fixed rules, plotting, for a population of UI
and REP,  the final (equilibrium) value of the relative cooperator density $\mu^{\scriptscriptstyle T}_{\scriptscriptstyle UI}$ and $\mu^{\scriptscriptstyle T}_{\scriptscriptstyle REP}$ as a function of $f_{\scriptscriptstyle UI}(0)$.
We indeed observe that the initial ratio of agent types has a significant effect on the final outcome of the simulations, although
this is not  so in
the case of UI (solid line): up to a value $f_{\scriptscriptstyle UI}(0)\sim0.75$ of the total initial fraction $f_{\scriptscriptstyle UI}(0)$
 the ratio $\mu^{\scriptscriptstyle T}_{\scriptscriptstyle UI}$ of UI cooperators is more or less constant, increasing slightly for
 larger values of $f_{\scriptscriptstyle UI}(0)\sim0.75$. In the case of REP agents (dashed line), we get another interesting result: The initial fraction of replicator agents $f_{\scriptscriptstyle REP}(0)=1-f_{\scriptscriptstyle UI}(0)$ is approximately proportional to the amount of replicator cooperators within the replicator population,  $\mu_{\scriptscriptstyle REP}$ being smaller (larger) than 0.5 whenever $f_{\scriptscriptstyle REP}(0)$ is smaller (larger) than 0.5. While this may look intuitive, it must be recalled that when the population
consists only of REP agents, the cooperators die out for any value of the temptation parameter
$b$. Regarding the time evolution of the experiment, the right panel of Fig. \ref{fig1} shows
two examples of the evolution of the normalized fraction $\mu_{\scriptscriptstyle UI}$ for
different values of the initial fraction of imitators $f_{\scriptscriptstyle UI}(0)$. There is
always an initial drop of $\mu_{\scriptscriptstyle UI}$ and a final relaxation to a value
$\mu^{\scriptscriptstyle T}$ that, as stated before, depends on $f_{\scriptscriptstyle
UI}(0)$.

\begin{figure}[h!]
\centering
\includegraphics[width=0.7\textwidth]{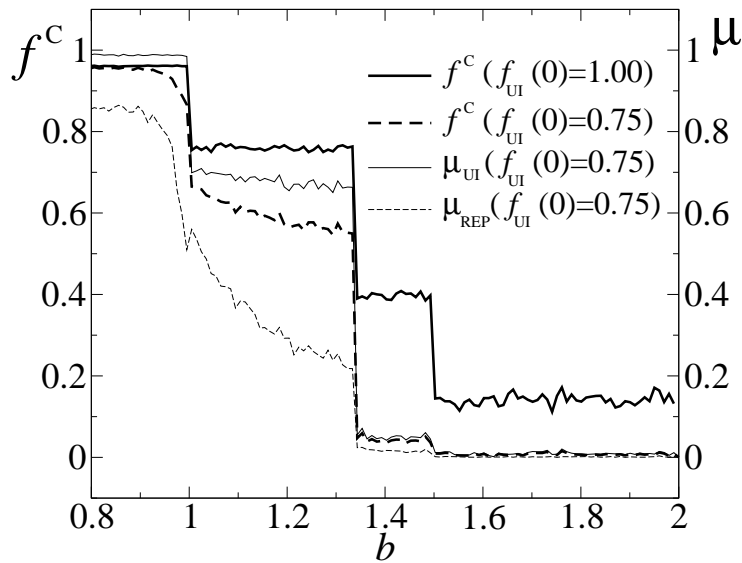}
\caption{
Density of cooperators for the UI vs. REP game with fixed dynamical rules. In thick solid lines, we present the case of a pure imitator population. In thick dashed lines, same settings but with $f_{\scriptscriptstyle UI}(0)=0.75$. In thin solid (dashed) line we show the normalised density of cooperator imitators (replicators) $\mu$. All quantities are averages of over 30 realisations.
}
\label{fig2}
\end{figure}

Having considered the effect of different populations of update rules,
let us now discuss the dependence on the temptation parameter $b$, and let us compare
the results for the homogeneous case (all agents equal) with the case where two different rules
are present. As a specific example, in
Fig. \ref{fig2} we plot, for the same game as before, the dependence of the total cooperator density as a function of the parameter $b$ for the homogeneous case $f_{\scriptscriptstyle UI}(0)=1$ and for a mixed case with fixed $f_{\scriptscriptstyle UI}(0)=0.75$.
For the homogeneous case, using the fact that unconditional imitation is a deterministic rule, one
can
show that $f_{\scriptscriptstyle UI}$ depends on $b$ in a step-like fashion, changing when $b$ is $\frac{4}{4}, \frac{4}{3}$ and $\frac{3}{2}$, in perfect agreement with the results of our simulations.
As we may see from the plot, this
functional dependence changes when there are replicators among the imitators.
We find that the effect of the replicators is to lower the amount of total cooperators for a the whole
 range of values of $b$ corresponding to the PD, e.g. $b \ge 1 $. In this respect, it is
interesting to look at how the fraction of cooperators changes relative to the population with
the same rule. This is also plotted in Fig. \ref{fig2}, by showing both
$\mu_{\scriptscriptstyle UI}$  and $\mu_{\scriptscriptstyle REP}$ as a function of $b$. We see
that, for $1 < b < \frac{3}{2}$, even though the total cooperator density is lower than in the
pure UI case, the cooperative imitators are a fraction larger than the total average, whereas
the replicators are less than the average. On the other hand, for $b>\frac{3}{2}$, defectors
take over the entire population, so  $\mu_{\scriptscriptstyle UI} = \mu_{\scriptscriptstyle
REP} =0$. This means that the level of cooperation attained for a pure UI population in the PD
suffers a considerable decrease or are totally suppressed by the presence of a minority of REP
agents (we have checked with populations as small as $f_{\scriptscriptstyle REP}=0.01$
obtaining similar results). We will come back to this issue when considering different
neighborhoods below.

\subsection{Variable dynamical rule}

In the previous subsection, we have reported that the presence of two different update rules in the
population may considerably change the behavior of the PD on a lattice, the general conclusion
being that the level of cooperation is lower than for the pure UI population. This result must be
kept in mind when analysing the outcome of allowing the update rules themselves to evolve, the
issue which we focus upon in the following.
When agents are allowed to switch their update rule when updating their strategies, then the fraction of agents of rules A and B, $f_{\scriptscriptstyle A}$, $f_{\scriptscriptstyle B}$, will generally change.
Indeed, in this case, we will see that, for certain values of the game parameters, a rule can completely overtake the system as the other one disappears, yielding different outcomes than the ones obtained with fixed update rules, or with just one update rule.

\begin{center}
  \begin{figure}[h!]
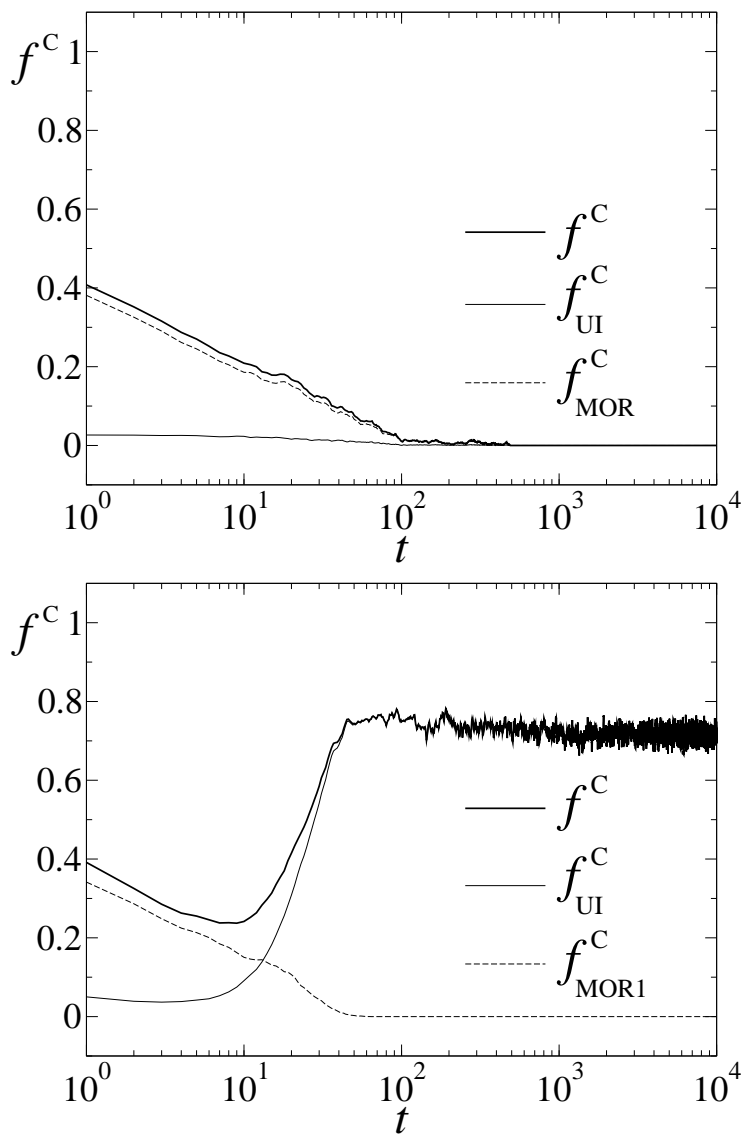

    \centering
    $\begin{array}{c}
      \includegraphics[width=0.7\textwidth]{figs/fig3aNEW.eps}\\
      \includegraphics[width=0.7\textwidth]{figs/fig3bNEW.eps}\\
    \end{array}$
    \caption{
      Top panel: The UI vs. MOR game with fixed rules for           $f_{\scriptscriptstyle UI}(0)=0.12$ and $b=1.20$. In solid line, the total cooperator                 density $f^{\scriptscriptstyle C}$. In thin solid (dashed) line the imitator (Moran) cooperator density. Bottom panel: same simulation as in top panel with update rule changing permitted. Lines have the same meaning as in top panel.
    }
    \label{fig3}
  \end{figure}
\end{center}

To begin with, let us present and discuss a specific example:
In Fig. \ref{fig3} we collect the results for the competition between UI agents versus MOR agents.
On the top panel we show the cooperator frequency evolution with fixed update rules, i.e., when
agents cannot change their update rules.
We clearly observe that the total cooperator fraction $f^{\scriptscriptstyle C}$ disappears at about $t\sim 300$.
Note that both UI and MOR cooperators disappear at about the same time as defectors take over the entire population, in agreement with our conclusion of the previous subsection that inhomogeneous agents
lead to lower levels of cooperation. Subsequently, let us consider
the bottom panel of Fig. \ref{fig3}, where we show a simulation under the same conditions with the only difference that in this case agents also copy their neighbour's update rule if they copy their (C or D) strategy. Opposite to the situations with fixed update rules,
not only the total fraction of cooperators does not disappear, but also we find that cooperators
end up forming about three quarters of the total population.
Interestingly, there is an initial decrease in the cooperation fraction, in agreement with the predictions
of the well-mixed/mean field calculation summarized in Sec.\ \ref{sec:meanfield} (here REP is
replaced by MOR, but the argument is very similar and applies as well). It's only at a later stage when
the effect of spatial correlations, namely the formation of clusters of cooperators, enters into play and
leads to an increase of cooperation \citep{Hauert-Viejo,Nowak-May,Carlos}.
Analysing the individual rules, we see that MOR cooperators disappear roughly at the same pace (actually, a little faster than in the fixed rule case), so all cooperators remaining are of the UI type.
Remarkably, UI agents completely replace MOR agents in spite of the fact that
the initial fraction of imitators in this simulation is $f_{\scriptscriptstyle UI}(0)=0.12$.

We could continue our discussion of specific examples but, to avoid a very lengthy description, let
us only mention that other interesting outcomes may appear depending on the game
parameters and update rule combinations.
For instance, it is possible to see a coexistence of cooperators of one rule with defectors of the other rule, as occurs for imitators versus Moran agents with $f_{\scriptscriptstyle UI}(t=0)<0.5$ and $b=1.05$.
However, as we are more interested in general conclusions than on a detailed classification of all
the possible outcomes, we will switch to a more general viewpoint in what follows.

\begin{figure}[h!]
\centering
\includegraphics[width=0.7\textwidth]{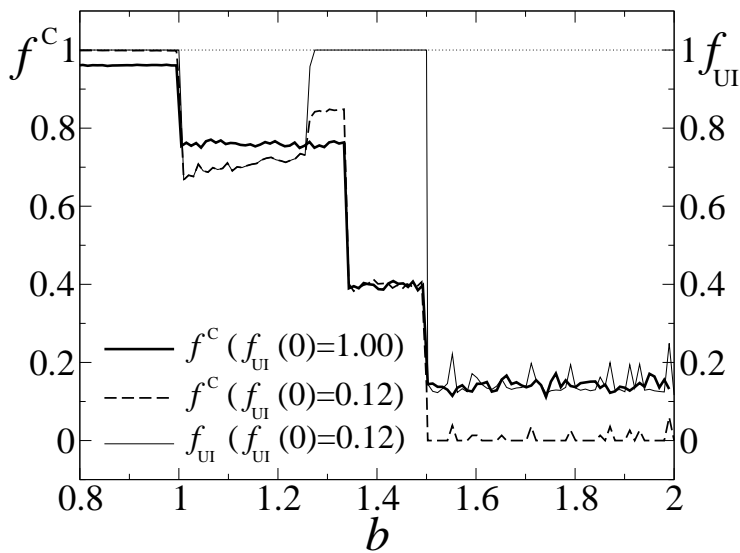}
\caption{
Density of cooperators $f^{\scriptscriptstyle C}$ and density of imitators $f_{\scriptscriptstyle UI}$ as a function of the temptation parameter $b$,
for the UI vs. MOR game. Simulation time is $T=10^4$. In thick solid line, we show $f^{\scriptscriptstyle C}$ for the case of a population composed by imitators only.
In dashed line, we present the same quantity for a mixed system with $f_{\scriptscriptstyle UI}(0)=0.12$.
In thin solid line, the final fraction of imitators (both cooperators and defectors). All quantities are averages of over 30 realisations.
}
\label{fig4}
\end{figure}

We will now discuss the dependence of the results on the temptation parameter $b$. In fact, as
agents are now allowed to change the dynamical rules, both the fraction of cooperators,
$f^{\scriptscriptstyle C}$, and the fraction of agents with rule A, $f_{\scriptscriptstyle
A}$, may change, so it is relevant to ask what is the dependence on the temptation parameter
$b$ for both of these quantities. In Fig. \ref{fig4} we show, as an example, our results for
the case of UI and MOR agents as a function of $b$. We see that depending on the value of $b$,
the interaction affects the final fraction of cooperators to different extents. Comparing with
the homogeneous UI case, we observe that there is a lower fraction of cooperators for the
mixed case when $1<b<1.25$ (the second bound is approximate) and for $b>\frac{3}{2}$. On the
other hand, the mixed case shows more cooperation when $b<1$ and in the interval
$1.25<b<\frac{4}{3}$. Finally, the two curves coincide for $\frac{4}{3}<b<\frac{3}{2}$. The
plot also depicts the final fraction of imitator agents $f_{\scriptscriptstyle UI}$ (cf.\
right axis) for the mixed case. Indeed, for most values of $b$, i.e. $b<\frac{3}{2}$, UI
agents have increased their frequency from the initial value $f_{\scriptscriptstyle
UI}(0)=0.12$), even becoming the \emph{only} rule ($f_{\scriptscriptstyle UI}=1$) for certain
regions ($b<1$ and $1.25<b<\frac{3}{2}$). Note that, for the cases $b<1$ and
$1.25<b<\frac{4}{3}$, the presence of Moran agents at early stages of the evolution makes the
equilibrium state to have more cooperation than the homogeneous UI case, even though Moran
agents finally dissapear completely. This constrasts with the case
$\frac{4}{3}<b<\frac{3}{2}$, where the cooperator frequency is the same as in the homogeneous
case.

\section{Extensions}

\subsection{Moore neighborhood}

\begin{figure}
\centering
\includegraphics[width=.9\textwidth,angle=0]{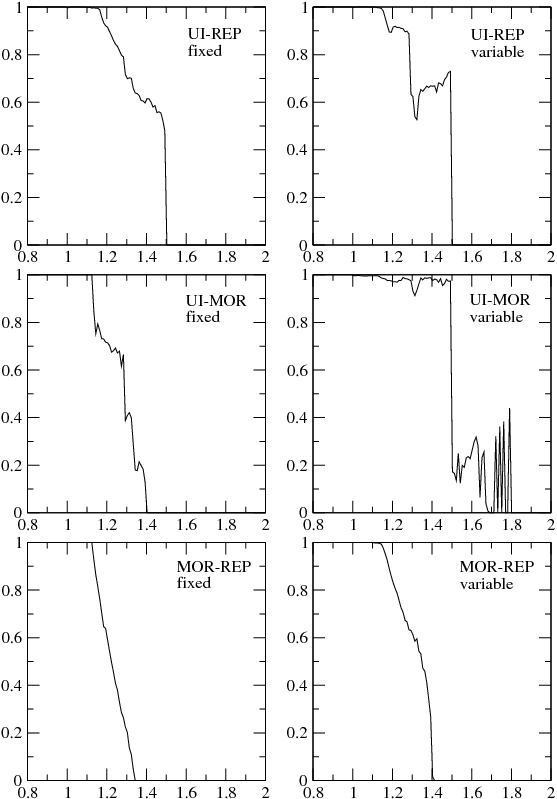}
\caption{ Mixed games with Moore neighbourhood and self-interaction
(8+1 neighbours), for fixed and variable rules and
$f^{\scriptscriptstyle C}(0)=0.50$. Initial rule composition is half
and half for all cases. } \label{g123}
\end{figure}

The results in Subsec.\ \ref{subsec:fixed} are an indication that
conclusions such as the promotion of cooperation on the PD on
lattices found by \cite{Nowak-May} may not be robust against the
presence of other types of update strategists, and therefore that
their applicability must be studied depending on the way individual
agents behave. However, those results have been obtained under the
restriction that players interact only with their nearest-neighbors
(von Neumann neighborhood), whereas Nowak and May included
next-nearest neighbor interactions (Moore neighborhood). To check
that the size of the neighborhood does not change our conclusions,
we have carried out simulations with the Moore neighborhood, the
results being depicted in Fig.\ \ref{g123}. In fact, we have
reproduced
 the simulations reported by \cite{Nowak-May}, which do not correspond exactly to a Moore neighborhood
 in so far as every player plays also with herself, i.e., it is an 8+1 neighborhood. As we may see from
 the example in the figure, we again find much lower levels of cooperation than in the pure imitator
 case, irrespective of whether the learning rules themselves evolve (right panels) or not (left panels).
 We note that this is not a general feature, as striking differences have been found between the
 presence and the absence of next-nearest-neighbor interactions, which makes our conclusion
 even more relevant as it is not trivial.
 We thus see that indeed the existence of different update rules in the population hinders cooperation
 on the square lattice.

\subsection{Asynchronous updating}

\begin{figure}
\centering
\includegraphics[width=.9\textwidth,angle=0]{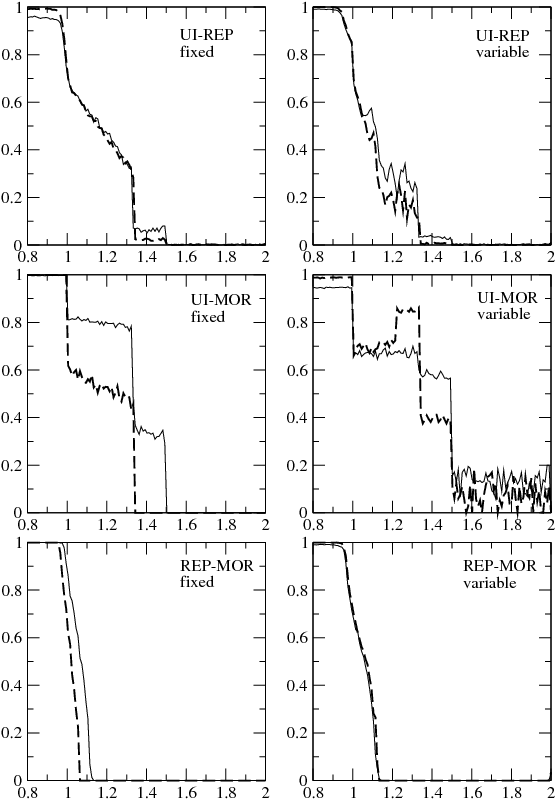}
\caption{Mixed games with synchronous (thick, dashed line) and
asynchrounous (thin, solid line) update, for fixed and variable
rules and $f^{\scriptscriptstyle C}(0)=0.50$. Initial rule
composition is half and half for all cases. } \label{copy-nocopy}
\end{figure}

Although it is beyond our purposes to go into an exhaustive study of asynchronicity, we think that
providing at least some examples of that case will increase the value and relevance of our results.
To this end, we have run some simulations in which every time step a single agent is chosen at
random and updates her strategy according to her current learning rule. Our results are summarized
in Fig.\ \ref{copy-nocopy} for fixed update rules (left) and changing update rules (right). We see
from the plot that although there are a few quantitative differences, particularly relevant for the
competition between UI and MOR with fixed update rules, the behavior is qualitatively the same
in the synchronous and the asynchronous updating schemes. It is thus clear that our conclusions
regarding the replacement of one rule by other are not an artifact of the synchronous update.

\section{Discussion}

\begin{figure}
\centering
\includegraphics[width=.95\textwidth,angle=0]{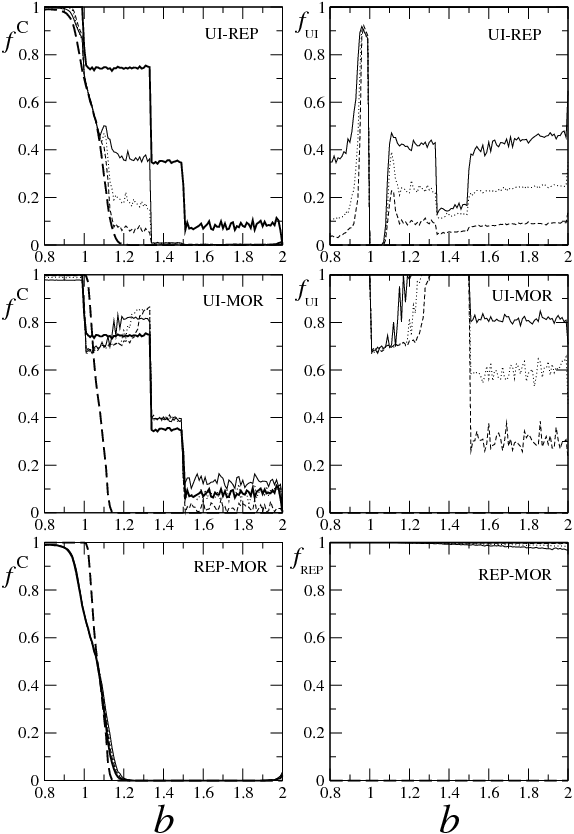}
\caption{ Asymptotic fraction of cooperators (left) and of players
with a given learning rule (right) for different initial
concentrations of that rule for the cases UI vs REP (top), UI vs MOR
(middle) and REP vs MOR (bottom). Thick solid lines, initial
condition with a 100\% of players of the first rule; thin lines
correspond to mixed populations: solid, 75\%, dotted, 50\% and
dashed, 25\%, always of the first rule. Finally, thick dashed lines
correspond to pure populations with a 100\% of the second rule. All quantities are averages of over 5 realisations.}
\label{nueva-global}
\end{figure}

As we have already said, our goal in this work is to extract general
conclusions, and therefore we will now proceed to summarize the main
results of our experiments. We simulated three possible
competitions, namely UI vs MOR, UI vs REP, and REP vs. MOR, for both
fixed and variable update rules, exploring the whole interval of
temptation values as well as initial conditions with different
composition for each case. Our results are summarized in Fig.\
\ref{nueva-global}, where we can see that, generally speaking, the
dominant update rule is REP, in the sense that in most situations it
dominates over the other rules even if its initial population is
small. This is particularly so in the case of REP vs MOR, for which
the presence of MOR agents in the final population is almost
negligible even if the initial population contains only a 25\% of
REP agents (see Fig.\ \ref{nueva-global} bottom). The REP rule also
dominates over the UI rule, although in this case UI does not go
fully extinct except for very small ranges of parameters, and, in
turn UI prevails over MOR, again without driving it to extinction.
This is also shown in the asymptotic cooperation levels (Fig.\
\ref{nueva-global}, left panels): when the population is a mixture
of REP with either UI or MOR, the dependence of the cooperation
level on $b$ is similar to that of a full REP population, with some
influence of the other rule (e.g., the abrupt drops in cooperation
at certain values of $b$ when the mixed population consists of REP
and UI agents). A remarkable feature is that, when the population is
a combination of UI and MOR agents, there is a range of values of
$b$, between 1.25 and 1.5 (and up to 2, depending on the initial
concentration of imitators), for which the cooperation level is
larger than for imitators, but for parameter values and populations
other than this case, the cooperation level is smaller than that
reached in a full UI population. This result suggests that the good
behavior observed on lattices and other networks homogeneous in
degree when the agents are of the UI type may not be very robust to
perturbations arising from agents learning with other rules. In any
event, it is clear that the lattices with mixed populations do
support cooperative states, which is a most important difference
with the well-mixed population case discussed in Sec.\
\ref{sec:meanfield}.

Regarding the prevalence of REP, we believe that it
is a consequence of its own dynamic mechanism, that has a built-in tendency to avoid changing its rule relative to the other dynamics tested in this work.
Indeed, if we compare the definitions of the three dynamics, we notice that UI agents always updates its strategy (and consequently its update rule), excepting when the other agent has less or equal payoff;
MOR agents
have a set of probabilities that will always trigger the rule updating, unless all probabilities are zero, as in the very particular case of zero payoff for all neighbours (for example, in a defector-only population).
Otherwise also the MOR agents update their rule with large probability, even
adopting another rule with less payoff, leading to the preservation of a rule that performed poorer;
and, finally,
a REP agent randomly chooses a neighbour, and then assigns a probability to adopt its rule
proportional to the payoff difference between both agents. For this rule
there is only one possibility for having certainty of the rule update, and this is the case where all neighbours are defectors fully surrounded by cooperators, and the updating agent is a cooperator, in turn fully surrounded by defectors (in a chessboard-like fashion). In this case we have certainty because of our normalization
factor for the probability, but it could even be the case that larger normalisation factors were used,
which would lead to a large but smaller than 1 probability to change.

\begin{figure}[h!]
\centering
\includegraphics[width=0.7\textwidth,angle=0]{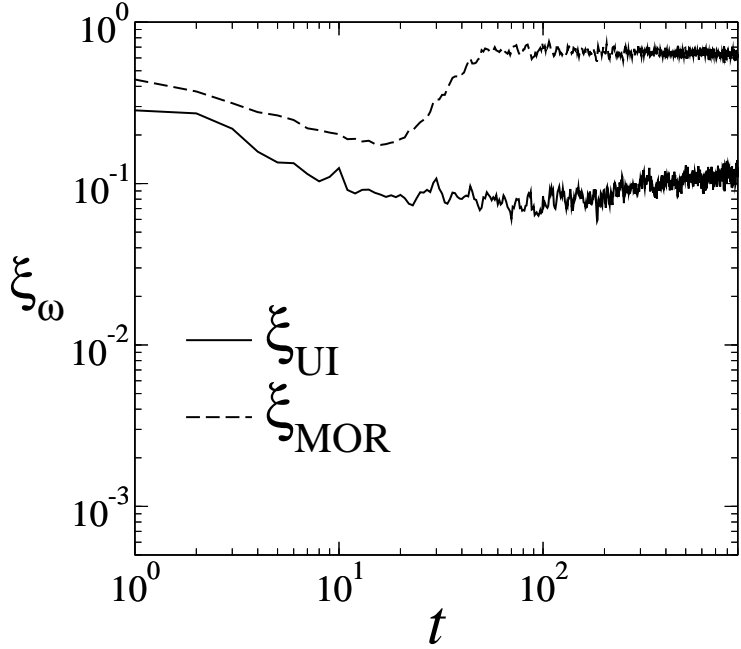}
\caption{
Rate of change of imitator $\xi_{\scriptscriptstyle UI}$ and Moran $\xi_{\scriptscriptstyle MOR}$ agents, relative to their own population. Parameters are the same as Fig. \ref{fig3} (bottom panel). Moran agents change most of the time at a higher rate than imitators (note the logarithmic scale in time).
}
\label{fig5}
\end{figure}

With this in mind, we monitored the fraction of agents of rule A (B) that update their rule
relative to the total population having rule A (B), in each generation, called $\xi_A$
($\xi_B$). As an illustration of the observed behaviours, in Fig. \ref{fig5} we plot this
quantity for the same case as Fig. \ref{fig3} (bottom panel), i.e., UI vs MOR agents. The
fraction of changing Moran agents rapidly becomes higher than that of imitators, and before
100 time steps is (and remains) much higher. We repeated this observation for many other cases
as well and, in general, our results confirm that, in most cases, REP agents change their
strategy and update rule much less frequently in comparison to the other dynamics. In some
cases, this difference may be an order of magnitude smaller for the replicator rule. On the
other hand the rule that updates the most is the Moran dynamics. Once again, this is
consistent with our results: Most times, MOR agents disappear completely when confronting REP
agents, and they survive in a restricted interval $1<b<1.1$ against UI, never being more than
half of the final population (see Fig. \ref{fig4}, thin solid line), whereas for
$\frac{4}{3}<b<\frac{3}{2}$ their presence is solely due to finite size effects. For UI vs
REP, imitators change their rule more frequently, and replicators appear to be systematically
the major part of the population.

\begin{center}
  \begin{figure}[h!]
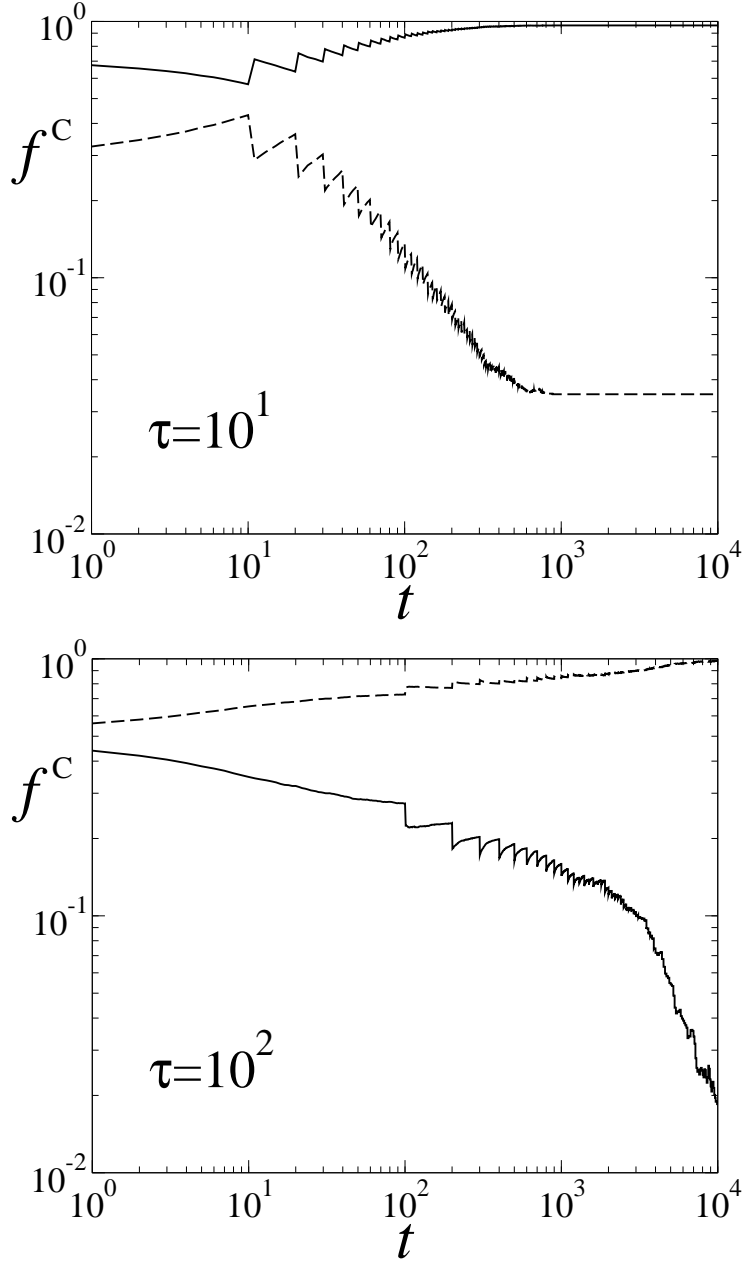

    \centering
    $\begin{array}{c}
      \includegraphics[width=0.7\textwidth]{figs/fig6a.eps}\\
      \includegraphics[width=0.7\textwidth]{figs/fig6b.eps}\\
    \end{array}$
    \caption{
      Normal replicator versus slow Moran agents, for different values of the parameter $\tau$. Temptation parameter is $b=1.2$ and $f_{\scriptscriptstyle REP}(0)=0.5$.
      Top panel: For $\tau=10$ evolution of the cooperator density. Replicators (solid line) outperform Moran agents (dashed line), as in the normal ($\tau=1$) case.
      Bottom panel: For $\tau=10^2$ the opposite happens.
    }
    \label{fig6}
  \end{figure}
\end{center}

In order to further investigate the hypothesis that the fraction of changing agents determine
their prevalence, we deviced the following experiment: We performed simulations with a similar
setup as before, but in this case one of the populations updates their strategies and rules
only at times $t=n \tau$, with $n=0, 1, 2\ldots$, being $\tau$ fixed (these would be ``slow''
agents). In this respect,  it is interesting to recall that a few recent works have proposed
that the existence of two types of agents, one of them with a smaller capability to transfer
its strategy, may promote cooperation \citep{joputa1,joputa2,joputa3}. Results of applying
this two-time scale setup to our simulation are collected in Fig. \ref{fig6}, where we see an
example for normal REP agents against slow MOR agents. These results confirm our hypothesis:
With $\tau=10^1$, replicators prevail as usual (top panel), whereas for a sufficiently high
update time, $\tau=10^2$, MOR agents prosper, and replicators tend to become extint (bottom
panel).

\begin{figure}[h!]
\centering
\includegraphics[width=0.7\textwidth,angle=0]{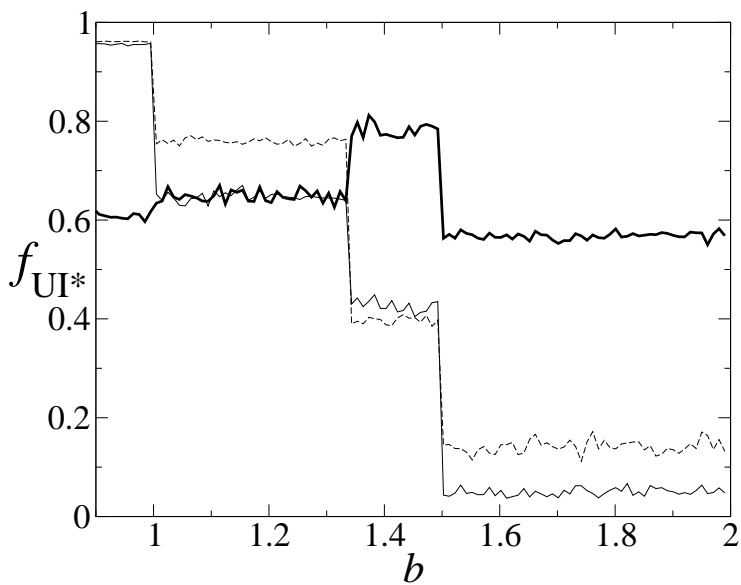}
\caption{
Fraction of slow imitator agents $f_{\scriptscriptstyle UI^*}$ (thick solid line) as a function of the temptation parameter $b$. For comparison,
we show also the total cooperation fraction $f^{\scriptscriptstyle C}$ (thin solid line) and the homogeneous (i.e., all agents with $\tau =1$)
cooperator frequency (thin dashed line). Initial fraction of slow agents is $f_{\scriptscriptstyle UI^*}(0)=0.50$ and $\tau_{\scriptscriptstyle UI^*}=10^2$.
}
\label{fig7}
\end{figure}

However, while this is an appealing mechanism, it cannot be the only responsible for the
prevalence of strategies. To show it, we considered the situation where agents of the same
kind are put together, the inhomogeneity being only that they can be ``normal'' or ``slow'',
i.e., that there is one fraction with a different $\tau$. In Fig. \ref{fig7} we show the
result of normal UI vs slow UI (see parameters in caption), where we observe that for all
values of the temptation parameter $b$, there is still a fraction of normal agents. If our
hypothesis above were the only reason for the survival of one strategy, the result should be
that slow UI would always take over the entire population, but this is clearly not the case.
Indeed, normal agents do not dissapear, and they only decrease in frequency noticeably (to
$f_{\scriptscriptstyle UI}\approx 0.2$) in the interval $\frac{4}{3}<b<\frac{3}{2}$. We
obtained similar results for a wide range of $\tau$, up to $\tau=10^3$. This indicates that,
at least for agents of the same type, the overall rule selection mechanism is not
straightforward and should be further studied, and we believe that the conclusions would also
carry over to the two update rule case. In addition, this makes clear that the two-learning
rule model is a  scenario that goes beyond that proposed by \cite{joputa1,joputa2} and
\cite{joputa3} and therefore deserves further attention to be completely understood.

A last point we want to remark from a general viewpoint is the
following. Within our simulation procedure, a population
composed only by defectors do not interact. Indeed, in this
case the only available payoff is $P=0$ independent of the
agent's update rule. REP or UI will not update their rule (they
do so only with more successful neighbours), and MOR agents
build a set of zero probabilities, thus keeping their rule as
well. Thus, irrespective of the rules, the system remains in a
``frozen'' state, with no further evolution. A similar
situation is found with a population composed only by
cooperative imitators and replicators. In this case, all agents
receive $R=1$ and thus have the same payoff. Therefore, there
is no rule update, and the system is, too, in a frozen state.
On the other hand, this is not the case when any of these rules
are set against the Moran rule, where evolution does occur, and
in most cases Moran players disappear. While all these are
direct consequences of the design of our simulation, we have
also found an interesting result regarding these locked
situations, where no evolution is possible, namely that the
presence of a small amount of a third type of agent may lead to
evolution and, moreover, to a completely different state.
Indeed, when these few agents can interact with at least one of
the other type of (otherwise frozen) players, the system as a
whole may evolve, even with the possibility of the extinction
of one or more species, that otherwise would be present. We
have not studied in detail this three update rule scenarios,
but we envisage that the dynamics will be much richer and
therefore deserves an in-depth analysis which is beyond the
scope of the present paper.

\section{Conclusions}

In this paper we have presented a first attempt to provide an evolutionary rationale for the
update rules used in spatially distributed models of emergence of cooperation. This is a key
aspect of these models in so far as the choice of update rule largely influences the
appearance and stabilisation of cooperation. In this context, our work must be viewed as a
proposal for a general mechanism that would allow modellers to decide upon the rule of
application in specific contexts. The main ingredient of this mechanism is the evolvability of
update rules according to the same scheme as the strategies themselves, i.e., when an agent
looks at her neighborhood and decides to copy the strategy of another agent, she also copies
the agent's update rule. In this respect, we want to stress that a related approach was
proposed by \cite{harley} as a rationale to explain how populations can learn the evolutionary
stable strategy. His results, which involve strategies with memory and referred to accumulated
payoff, relate the stability of learning rules to their ability to take the population to the
evolutionary stable strategy. Although our proposal here is quite different, in particular
because our focus is not reaching an equilibrium but rather letting the system evolve at will,
it is only fair to acknowledge Harley's pioneering work in proposing the evolvability of
strategies. On the other hand, our model, much as Harley's procedure,  can be interpreted as
learning in social contexts, and hence it has in turn a much more biological character than
endogenous learning rules such as those introduced by \cite{kirch}, that, to our knowledge, is
the only other study where a evolutionary origin of learning rules has been explored. Note,
however, that \cite{kirch} considered a variety of games, while here we focus on the
Prisoner's Dilemma in view of its applicability in a number of social, biological and even
biochemical systems. Finally, it is worth mentioning recent work by  \cite{private}, where
Darwinian selection is applied to a one-parameter stochastic update rule (similar to those
used by \cite{joputa1,joputa2}), leading to the selection of a specific value of the
parameter. This work is not related to ours in the update rule they use but the spirit is
quite the same.

Beyond this general statement of the importance of the idea of evolvability of 
learning rules and its role to decide which ones should be used, we have reached several
important conclusions of our work, that to our knowledge have never been 
reported elsewhere. In fact, their relevance arises not only for their own sake but also because
they affect to two of the most often employed learning rules (UI and REP, see \cite{Szabo}) and 
because among the three rules we consider we cover the options local-global and 
deterministic-stochastic. Our main findings  can thus be summarized as follows:
\begin{itemize}
\item A well-mixed population playing the Prisoner's Dilemma evolves to full defection 
when individuals have two learning strategies, even if these learning strategies can in
turn evolve. Section 3 provides an example for replicators vs imitators but other combinations
can be worked out in a similar manner. 
\item When the population is placed on a square lattice, the existence of individuals with
two different, permanent learning rules leads to dramatic changes with respect to the 
separate cases of the two strategies. Using again the example of replicators vs imitators 
(Subsection 4.1),
a small proportion of replicators may lead to the breakdown of cooperation generically
observed for imitators  \citep{Nowak-May}. 
\item Evolvability of learning rules has crucial implications on the outcome of evolutionary
game dynamics. Thus, Subsection 4.2 shows that 
replicator displaces imitators leading to a rapid decreasing
of cooperation. In turn, imitators take over the fraction of global but stochastic 
imitators (MOR) and lead to
a promotion of the cooperative behavior. The phenomenon, however, is not trivial, and 
regions where evolution leads to an outcome opposite to what is expected in general 
are also observed (e.g., Fig.\ 4, $1.25<b<4/3$).
\item This competition process takes place on lattices under different environments. Section
5 shows that our results are robust against changes on the neighborhood considered (Moore 
instead of von Neumann) and against asynchronous dynamics. This indicates that our 
conclusions have a large degree of generality.
\item We have
shown that, in general, rules that lead to less frequent strategy changes (and consequently to
their own update in our model) tend to be selected, although our simulations also prove that 
this is not the whole story (Section 6).
\item Finally, as a general conclusion, our work makes it clear that modelling the emergence
of cooperation in the Prisoner's Dilemma on networks must go beyond the work done so far
in terms of populations with a single, constant in time learning strategy. While by no means
have we considered all possible rules or all possible networks, our simulations provide 
well-established evidence that the presence of different learning rules and their own evolution
may lead to unexpected phenomena, sometimes opposite to the observations available so 
far on unique-learning dynamics. Games with even more delicate equilibria structure such
as Snowdrift or Stag-Hunt are likely to be affected in yet a stronger manner. 
\end{itemize}

As a closing remark, we want to stress that, while we acknowledge the limited scope 
of the present report, 
we believe that our study opens the way to a much more complete analysis of the evolution of
the update rules. It is clear that the present work asks for further research, regarding,
e.g., the case in which three different update rules are present simultaneously. As we have
advanced above, in this scenario much more complicated process may appear as the third rule
helps other resist invasion by a dominant one. On the other hand, we have by no means
exhausted the possible update rules, and a more thorough simulation program which would
include more deterministic local rules is needed, in order to determine whether or not local
rules invade global rules, or whether stochastic is better than deterministic. A specific case
of relevance in social networks is that of reinforcement learning \citep{wang}, which is much
more complex than the rules considered here and might provide an interesting first step to
extend our results to more realistic situations. It would also be interesting to consider
different time scales for the update of strategies and the update of learning rules. Thus, it
is conceivable that an agent only changes her update rule after copying the strategy of the
same neighbour a certain number of times. Mutations in the learning rule can also be included,
and even heterogeneity in the details of the rule (for instance, different proportionality
factors in the REP rule). These and other extensions of the present work should include in
addition the comparison of the results of different games beyond the PD. We envisage that such
a programme would be extremely useful for clarifying the big picture of evolutionary game
theory on graphs and its applications.

\section*{Acknowledgments}

The authors are grateful to Carlos P\'erez Roca and Esteban Moro for helpful
and interesting discussions, and to Jos\'e A.\ Cuesta for his help with the well-mixed
population discussion. We are also grateful to Manfred Milinski for pointing out
the work by \cite{harley} to us.

This work was supported by Ministerio de Educaci\'on y Ciencia (Spain) under grant MOSAICO and by Comunidad de Madrid (Spain) under grant
SIMUMAT-CM.

\end{document}